\documentclass[aps,pra,onecolumn,11pt,superscriptaddress]{revtex4-2}
\usepackage{times}
\usepackage{xfrac}

\usepackage{amsmath}
\usepackage{amsfonts}
\usepackage{amssymb,bbm}
\usepackage[caption=false]{subfig}
\usepackage{color}
\usepackage{float}
\usepackage[T1]{fontenc}
\usepackage[utf8x]{inputenc}
\usepackage[english]{babel}
\usepackage{lmodern,braket}
\usepackage{bbold}
\usepackage{graphicx}
\usepackage{bm,bbm}
\usepackage{tikz-cd} 
\usepackage{tikz}
\usetikzlibrary{positioning}
\usepackage{adjustbox}
\usepackage{enumerate}
\usepackage[colorlinks=true,citecolor=blue,linkcolor=blue,urlcolor=blue]{hyperref}

\def\<{\langle}
\def\>{\rangle}

\newcommand{\hs}{\mathcal{L}_2(\mathcal{H}_S)}
\newcommand{\he}{\mathcal{L}_2(\mathcal{H}_E)}
\newcommand{\hse}{\mathcal{L}_2(\mathcal{H}_S\otimes \mathcal{H}_E)}

\newcommand\opone{\leavevmode\hbox{\small1\kern-3.8pt\normalsize1}}

\DeclareMathOperator{\Tr}{Tr}

\usepackage{url}
\usepackage{subfig}
\usepackage{xcolor}
\usepackage[normalem]{ulem}
\newcommand{\stkout}[1]{\ifmmode\text{\sout{\ensuremath{#1}}}\else\sout{#1}\fi}

\begin{document}

\title{On the use of total state decompositions for the study of reduced dynamics}

\author{Andrea Smirne}
\email{andrea.smirne@unimi.it}
\affiliation{Dipartimento di Fisica ``Aldo Pontremoli'', Universit\`a degli Studi di Milano, via Celoria 16, 20133 Milan, Italy}
\affiliation{Istituto Nazionale di Fisica Nucleare, Sezione di Milano, via Celoria 16, 20133 Milan, Italy}
\author{Nina Megier}
\email{nina.megier@mi.infn.it}
\affiliation{Dipartimento di Fisica ``Aldo Pontremoli'', Universit\`a degli Studi di Milano, via Celoria 16, 20133 Milan, Italy}
\affiliation{Istituto Nazionale di Fisica Nucleare, Sezione di Milano, via Celoria 16, 20133 Milan, Italy}
\affiliation{International Centre for Theory of Quantum Technologies (ICTQT), University of Gdańsk, 80-308 Gdańsk, Poland}
\author{Bassano Vacchini}
\email{bassano.vacchini@mi.infn.it}
\affiliation{Dipartimento di Fisica ``Aldo Pontremoli'', Universit\`a degli Studi di Milano, via Celoria 16, 20133 Milan, Italy}
\affiliation{Istituto Nazionale di Fisica Nucleare, Sezione di Milano, via Celoria 16, 20133 Milan, Italy}

\begin{abstract}
     The description of the dynamics of an open quantum system in the presence of initial correlations with the
     environment needs different mathematical tools than the standard approach to reduced dynamics, which is based on
     the use of a time-dependent completely positive trace preserving (CPTP) map.
     Here, we take into account an approach that is based on a decomposition of any possibly correlated
     bipartite state as a conical combination involving statistical operators on the environment and
     general linear operators on the system, which allows one to fix the reduced-system evolution
     via a finite set of time-dependent CPTP maps. In particular,
     we show that such a decomposition always exists, also for infinite dimensional Hilbert spaces,
     and that the number of resulting CPTP maps is bounded by the Schmidt rank of the initial global state.
     We further investigate the case where the CPTP maps are semigroups with generators in the Gorini-Kossakowski-Lindblad-Sudarshan
     form; for two simple qubit models, we identify the positivity domain defined by the initial states
     that are mapped into proper states at any time of the evolution fixed by the CPTP semigroups.
\end{abstract}

\maketitle

\section{Introduction}
\label{sec:intro}
The dynamics of open quantum systems in the presence of initial correlations
with the environment has been attracting a renewed interest, driven by both fundamental questions
and the need for a realistic description of concrete physical systems \cite{Grabert1988,Pechukas1994,Alicki1995,Lindblad1996,Buzek2001,Myohanen2008,Pomyalov2010,Modi2012,Vacchini2016a,Schmid2019,Majeed2019,Merkli2022}.
Relaxing the assumption that the open system and the environment are uncorrelated
at the initial time, the very existence of a reduced dynamics defined in terms 
of a time-dependent map acting on the whole set of open-system states is no longer guaranteed.
Most of the recent theoretical literature on the subject has thus been focused on how to extend
the notion of dynamical maps to the scenario with initial system-environment correlations, 
resorting to the identification of a restricted domain of allowed initial states, 
and possibly preserving the notion of complete positivity in such a scenario \cite{Jordan2004,Jordan2006,Rosario2008,Shabani2009,Brodutch2013,Buscemi2014,Dominy2016}.

Very recently \cite{Silva2019}, a different approach has been introduced, which, relying on the frame decomposition of the bipartite initial state, leads to the definition of a set of completely positive trace preserving (CPTP) time-dependent maps
that describe the evolution of a precisely identified set of initial states. 
More specifically, the exploited decomposition relies on the introduction 
of a positive frame on the set of the open-system Hilbert-Schmidt operators
and, provided that such a positive frame exists, it allows one to deal with fully general initial conditions.
The number of involved CPTP maps depends on the chosen initial state and it is anyway bounded
by the square of the dimensionality of the open system. 
This approach has been used to devise an efficient control of general open-system evolutions \cite{Chalermpusitarak2021a} and to
characterize multipartite photonic systems \cite{Raja2020};
furthermore, it has been combined with perturbative projection-operator techniques \cite{Breuer2002}
to derive an uncoupled system of homogeneous master equations that can be applied
under general initial conditions \cite{Trevisan2021a}.

In this paper, we introduce two different positive decompositions, one starting from the generalized Pauli matrices 
for a generic finite dimension, and one directly built on any orthonormal basis of a possibly infinite dimensional Hilbert space.
The latter, in particular, explicitly shows that the description of the open-system evolution in the presence of initial
correlations via a set of time-dependent CPTP maps can be defined in full generality, for any initial global
state and for any dimension of the open-system Hilbert space.
Moreover, in the finite dimensional case, we prove that the number of needed CPTP maps always coincides 
with the Schmidt rank of the initial global state, both for pure and mixed states. 
This clarifies in a quantitative way the enhanced complexity needed to describe open-system dynamics when moving
from an initial product state, where one CPTP map is enough, to initially correlated states. In the latter case, CPTP maps can still be
used, but at the price of increasing their number according to the Schmidt rank of the initial global state, which will be indeed
strictly larger than one in the presence of correlations.
Finally, in the last part of the paper, we have considered the case in which the CPTP maps
are semigroups with generators in the Gorini-Kossakowski-Lindblad-Sudarshan (GKLS) form. 
For the case of qubit Pauli channels, we have derived and investigated an explicit condition that defines
the set of initial open-system states 
that are mapped into positive states at any subsequent time of the evolution, 
when each of the semigroups is applied
to a different term of the conical decomposition yielding the initial reduced state.

The rest of the paper is organized as follows. 
In Sec.\ref{sec:one-sided-positive}, we briefly recall the framework
of open-system dynamics in the presence of initial system-environment correlations
and how it can be described through a family of time-dependent CPTP maps
via the approach introduced in \cite{Silva2019}, and we further recall the elements of the theory of frames
that we will exploit in the following.
In Sec.\ref{sec:eacc}, we show that this approach can be extended to infinite-dimensional Hilbert spaces of the open-system,
and we prove that in the finite dimensional case the number of required CPTP maps can be linked in full generality to the Schmidt rank
of the initial global state.
In Sec.\ref{sec:csc}, we specialize our analysis to the case were the CPTP maps are semigroups and
we give a detailed characterization of initial reduced states compatible with such a choice,
in the case of qubit Pauli channels.
Finally, in Sec.\ref{sec:con}, the conclusions of our work are presented.

\section{One-sided positive decomposition}
\label{sec:one-sided-positive}

\subsection{Open quantum system dynamics via completely positive maps}
\label{sec:OQSvCP}
We consider the standard framework to describe the dynamics of an open quantum system \cite{Breuer2002}.
We have a bipartite global system, consisting of the open system $S$, 
which is associated with the Hilbert space $\mathcal{H}_S$,
and the environment $E$, associated with $\mathcal{H}_E$, 
so that the global system is associated with the tensor product Hilbert space 
$\mathcal{H}_S \otimes \mathcal{H}_E$. Moreover, the global system is assumed to evolve 
unitarily, i.e., the joint system-environment evolution is fixed by the unitary operators 
$U(t)$ (here and in the following, $t_0=0$ is the initial time).
Let $\mathcal{S}(\mathcal{H})$ denote the set of statistical operators, i.e. positive linear operators with unit trace, on $\mathcal{H}$.
The reduced open-system state $\rho_S(t)$
can be written at any time $t$ as a map from 
$\mathcal{S}\left(\mathcal{H}_{S}\otimes \mathcal{H}_{E}\right)$ 
to $\mathcal{S}(\mathcal{H}_S)$,
\begin{equation}\label{eq:rst}
\rho_S(t) = \Tr_E \left[U(t)\rho_{SE}U^{\dagger}(t)\right],
\end{equation}
where $\mbox{Tr}_E$ is the partial trace on the environmental degrees of freedom.
The map defined in Eq.(\ref{eq:rst}) is CPTP, but its domain involves the whole
$\mathcal{S}\left(\mathcal{H}_{S}\otimes \mathcal{H}_{E}\right)$, while 
when dealing
with the evolution of an open quantum system, one would like to 
describe the dynamics via maps defined on $\mathcal{S}(\mathcal{H}_S)$ only. 

This goal is achieved in \cite{Silva2019}, relying on the
decomposition of the bipartite statistical operator
$\rho_{SE}\in \mathcal{S}\left(\mathcal{H}_{S}\otimes \mathcal{H}_{E}\right)$ as
\begin{equation}\label{eq:opd}
\rho_{SE}= \sum_{\alpha=1}^{\mathfrak{N}} \omega_{\alpha} D_{\alpha}\otimes \rho_{\alpha},
\end{equation}
where $\rho_{\alpha} \in \mathcal{S}(\mathcal{H}_E)$ and
$\omega_\alpha \geq 0$, while the $D_{\alpha}$ are operators 
within the set $\hs$ of Hilbert-Schmidt operators on
$\mathcal{H}_S$, i.e., the trace of the square of their absolute value is finite; 
note that these operators are indeed not necessarily positive.
We will call the representation of $\rho_{SE}$ in Eq.(\ref{eq:opd}) \textit{one-sided positive decomposition} (OPD),
to stress that positive operators on the environmental side are needed; 
we will further call \textit{cost}
of the OPD the minimum number $\mathfrak{N}$ of terms with which the sum in Eq.(\ref{eq:opd}) can be expressed. 
If the $D_{\alpha}$ are positive operators, $\rho_{SE}$ in Eq.(\ref{eq:opd})
is a separable state \cite{Bengtsson2006}; if, in addition, the $D_{\alpha}$ or the $\rho_{\alpha}$ or both
are given by a family of orthogonal projections, $\rho_{SE}$ is a zero discord state \cite{Ollivier2001,Henderson2001,Modi2012}. 
On the other hand, let us stress once
more that general bipartite states, including any kind of classical or quantum correlations, 
can be decomposed via (infinitely many) OPDs.

From the point of view of the description of open-system dynamics, the key advantage
of the OPD is that, starting from it, the reduced state at time $t$ can always be obtained 
by means of a family of time-dependent CPTP maps defined on operators
acting on $\mathcal{H}_S$ only. Replacing Eq.(\ref{eq:opd})
into Eq.(\ref{eq:rst}), we have in fact
\begin{align}\label{eq:opdt}
	\rho_S(t)&=\sum_{\alpha=1}^{\mathfrak{N}} \omega_{\alpha} \Phi_{\alpha}(t)[D_{\alpha}],
\end{align}
where 
\begin{align}\label{eq:phialpha}
	\Phi_{\alpha}(t): \hs&\to \hs \\
							A &\mapsto \Phi_{\alpha}(t)[A]=\Tr_E[U(t)A\otimes \rho_{\alpha}U(t)^{\dagger}]. \notag
\end{align}
The initial reduced state
\begin{equation}\label{eq:inits}
\rho_S = \sum_{\alpha=1}^{\mathfrak{N}} \omega_{\alpha} D_{\alpha}
\end{equation}
is mapped into the reduced state at time $t$
by the $\mathfrak{N}$ CPTP maps $\left\{\Phi_{\alpha}(t)\right\}_{1,\ldots, \mathfrak{N}}$ on $\hs$.
An initial product state corresponds to the case $\mathfrak{N}=1$, which directly reduces to the usual
description in terms of a single CPTP map \cite{Breuer2002,Rivas2014}.
The presence of initial correlations implies that generally
$\mathfrak{N}>1$ CPTP maps are needed; on the other hand, the same
set $\left\{\Phi_{\alpha}(t)\right\}_{1,\ldots, \mathfrak{N}}$ of maps can be used 
for all the states connected by
any local operation on $S$ \cite{Silva2019}. 

\subsection{Theory of frames: reconstruction formula and positive frames}
\label{sec:frame-theory-posit}
Here, we briefly present the main features of the theory of frames that will be relevant for our purposes,
directly referring to the set $\hs$; we will further
restrict to frames consisting of a countable set of operators, but, indeed, extensions to uncountable
sets are possible. For a more general treatment of frame theory, the interested reader 
is referred to \cite{Ali2000,Renes2004a}.

Consider the Hilbert space of Hilbert-Schmidt operators $\hs$ equipped with the scalar product
\begin{equation}\label{eq:hssc}
\left(A, B \right) = \mbox{Tr}_S \left[A^\dag B \right], \quad A, B \in \hs,
\end{equation}
where $\mbox{Tr}_S$ denotes the trace over $\mathcal{H}_S$,
and a family of operators $\left\{F_\alpha\right\}_{\alpha \in I}$, where each 
$F_\alpha \in \hs$ and the index $\alpha$ takes values in the countable
set $I$, such that 
\begin{equation}\label{eq:cond}
\sum_{\alpha \in I} |\left(F_\alpha, A\right)|^2< \infty
\end{equation}
for every $A \in \hs$.
The associated map on $\hs$
\begin{align}\label{eq:xi}
	\Xi: \hs&\to \hs \\
		A &\mapsto \Xi[A]=\sum_{\alpha \in I} \left(F_\alpha, A\right) F_\alpha \notag
\end{align}
is called \emph{frame map};
the operators $F_\alpha$ are indeed not necessarily orthogonal, nor linearly independent.
Now, one says that the family of operators $\left\{F_\alpha\right\}_{\alpha \in I}$
is a \emph{frame} of $\hs$ whenever Eq.(\ref{eq:cond}) holds and the corresponding frame map $\Xi$
satisfies the following lower and upper bound:
\begin{equation}\label{eq:frcond}
m \|A\|_2 \leq (A, \Xi[A]) \leq M \|A\|_2, \quad \forall A\in \hs,
\end{equation}
for some $0<m\leq M< \infty$, where $\|A\|_2= \mbox{Tr}_S \left[A^\dag A\right]$ is the Hilbert-Schmidt norm induced by 
the scalar product in Eq.(\ref{eq:hssc});
in particular, the lower bound with $m$ strictly larger than zero implies that 
$\left\{F_\alpha\right\}_{\alpha \in I}$ spans $\hs$.
On the whole, the definition of frame is equivalent to the requirement that $\Xi$ is bounded and invertible,
with bounded inverse $\Xi^{-1}$. Using such properties, so that we can write
$$
A = \Xi^{-1}\left[\Xi[A]\right] =\sum_{\alpha \in I} \left(F_\alpha, A\right) \Xi^{-1}\left[F_\alpha\right], 
$$ 
we decompose any Hilbert-Schmidt
operator according to
\begin{equation}\label{eq:dec}
A = \sum_{\alpha \in I} (F_\alpha, A) \widetilde{F}_{\alpha}, \quad A\in\hs,
\end{equation}
where we defined
\begin{equation}\label{eq:indu}
\widetilde{F}_{\alpha}= \Xi^{-1}[F_\alpha].
\end{equation}
Importantly, Eq.(\ref{eq:dec}) 
allows us to reconstruct any operator in $\hs$ in terms of the elements of the chosen frame. Indeed, this is a generalization
of the reconstruction formula of any vector of a Hilbert space by means of an orthonormal basis. In fact, orthonormal bases are a special case of frames, for which $\Xi$ corresponds to the identity
map and thus Eq.(\ref{eq:frcond}) holds with $m=M=1$\footnote{Whenever Eq.(\ref{eq:frcond}) holds with $m=M=1$, the frame $\left\{F_\alpha\right\}_{\alpha \in I}$ is called a Parseval frame, as Eq.(\ref{eq:dec}) reduces to $A = \sum_{\alpha \in I} (F_\alpha, A) F_{\alpha}$; interestingly, there are Parseval frames that are \emph{not} orthonormal bases.}.

It is easy to see that the family $\left\{\widetilde{F}_\alpha\right\}_{\alpha \in I}$
is itself a frame, whose frame map coincides with $\Xi^{-1}$,
so that the corresponding reconstruction formula reads
$$
A = \sum_{\alpha \in I} (\widetilde{F}_\alpha, A) F_{\alpha}, \quad A\in\hs;
$$ 
$\left\{\widetilde{F}_\alpha\right\}_{\alpha \in I}$ is called the \emph{canonical dual frame}
of $\left\{F_\alpha\right\}_{\alpha \in I}$. More in general, a dual frame of 
$\left\{F_\alpha\right\}_{\alpha \in I}$ is a frame $\left\{D_\alpha\right\}_{\alpha \in I}$, such
that for any operator $A$ the following reconstruction formula holds:
\begin{equation}\label{eq:dec2}
A = \sum_{\alpha \in I} (F_\alpha, A) D_{\alpha} = \sum_{\alpha \in I} (D_\alpha, A) F_{\alpha}, \quad A\in\hs;
\end{equation}
as seen, every frame has at least one dual frame, which is the canonical dual frame.

We can now move back to the bipartite setting that is of interest for us, so that, besides the open system $S$, we also consider the environment $E$ and we deal with Hilbert-Schmidt operators on the tensor product $\mathcal{H}_S \otimes \mathcal{H}_E$, i.e.,
$O_{SE} \in \hse$.
Given a frame of operators referred to the open system, $\left\{F_\alpha\right\}_{\alpha \in I}$, and
a dual frame $\left\{D_\alpha\right\}_{\alpha \in I}$, with $F_\alpha, D_\alpha \in \hs$,
together with a frame of environmental operators, $\left\{X_\beta\right\}_{\beta \in J}$, along with 
a dual frame $\left\{Y_\beta\right\}_{\beta \in J}$, with $X_\beta, Y_\beta \in \he$, one can readily see that $\left\{F_\alpha \otimes X_\beta\right\}_{\alpha \in I, \beta\in J}$ and 
$\left\{D_\alpha \otimes Y_\beta\right\}_{\alpha \in I, \beta\in J}$ provide us with a frame of $\hse$
and its dual.
We can thus apply the reconstruction formula to any system-environment Hilbert-Schmidt operator,
\begin{align}
O_{SE} &= \sum_{\alpha \in I, \beta\in J} (F_\alpha \otimes X_\beta, O_{SE}) D_{\alpha} \otimes Y_\beta \nonumber\\
&= \sum_{\alpha \in I, \beta\in J} \mbox{Tr}_{SE}\left[(F^\dag_\alpha \otimes X^\dag_\beta) O_{SE}\right] D_{\alpha} \otimes Y_\beta \nonumber\\
&=\sum_{\alpha \in I} D_{\alpha} \otimes \left\{\sum_{\beta\in J}\mbox{Tr}_E \left[\mbox{Tr}_S\left[(F^\dag_\alpha\otimes\mathbb{1}_E) O_{SE}\right]X^\dag_\beta\right]Y_\beta\right\} \nonumber,
\end{align}
where in the second identity we used the definition of the Hilbert-Schmidt scalar product of $\hse$,
as in Eq.(\ref{eq:hssc}) but where the trace is now taken over all the global $S-E$ degrees of freedom,
while in the third identity we used that 
$\mbox{Tr}_{SE}\left[(F^\dag_\alpha \otimes X^\dag_\beta) O_{SE}\right] = \mbox{Tr}_E \left[\mbox{Tr}_S\left[(F^\dag_\alpha\otimes\mathbb{1}_E) O_{SE}\right]X^\dag_\beta\right]$, with $\mathbb{1}_E$ the identity operator on $\mathcal{H}_E$. Hence, noting that the expression in the curly brackets in the last
line is simply the reconstruction formula for the environmental operator 
$\Tr_S\left[(F^\dag_\alpha\otimes\mathbb{1}_E) O_{SE}\right]$ on the frame 
$\left\{X_\beta\right\}_{\beta \in J}$ and its dual $\left\{Y_\beta\right\}_{\beta \in J}$, we 
conclude that any global $S-E$ Hilbert-Schmidt operator $O_{SE}\in\hse$ can be decomposed as
\begin{equation}\label{eq:decbip}
O_{SE} = \sum_{\alpha \in I}  
D_{\alpha} \otimes \Tr_S\left[(F^\dag_\alpha\otimes\mathbbm{1}_E) O_{SE}\right];
\end{equation}
in other terms, a frame of open-system operators naturally induces a decomposition formula
for the global operators.

Comparing Eq.(\ref{eq:decbip}) with Eq.(\ref{eq:opd}), we see that in order to define 
an OPD we have to guarantee that when we focus on system-environment states, 
$\rho_{SE}\in \mathcal{S}(\mathcal{H}_S \otimes \mathcal{H}_E) \subset \hse$,
the decomposition can be expressed in terms of environmental statistical operators and positive coefficients. Indeed, this is the case if we have a \emph{positive frame}, i.e., a frame 
$\left\{F_\alpha\right\}_{\alpha \in I}$ made up of positive operators, $F_\alpha \geq 0$ for any $\alpha$,
which implies that $ \mbox{Tr}_S\left[(F^\dag_\alpha\otimes\mathbb{1}_E) \rho_{SE}\right] =\mbox{Tr}_S\left[(F_\alpha\otimes\mathbb{1}_E) \rho_{SE}\right] \geq 0$.
In fact, introducing the trace of these operators
\begin{equation}\label{eq:oma1}
\omega_\alpha = \mbox{Tr}_E\left[\mbox{Tr}_S\left[(F^\dag_\alpha\otimes\mathbb{1}_E) \rho_{SE}\right] \right] \geq 0
\end{equation}
and defining the environmental statistical operators $\rho_\alpha \in \mathcal{S}(\mathcal{H}_E)$
via the assignment
\begin{equation}\label{eq:omarhoa}
\omega_\alpha \rho_\alpha =  \mbox{Tr}_S\left[(F^\dag_\alpha\otimes\mathbb{1}_E) \rho_{SE}\right] 
\end{equation}
(so that $\rho_\alpha$ is arbitrary when 
$\mbox{Tr}_S\left[(F^\dag_\alpha\otimes\mathbb{1}_E) \rho_{SE}\right]=0$), we end up with 
an OPD. 
Furthermore, we note that the coefficients $\omega_\alpha$ only depend on the reduced
system state $\rho_S =\mbox{Tr}_E\left[\rho_{SE}\right]$, since they are equivalently expressed by
\begin{equation}\label{eq:oma2}
\omega_\alpha =  \mbox{Tr}_S\left[F^\dag_\alpha \rho_{S}\right]. 
\end{equation}

In the next section, we present two different explicit choices of positive operators 
allowing us to express every initial global state
with an OPD, but before that 
let us note the following. Once we have assigned a certain initial global state $\rho_{SE}$,
there is a (slightly) weaker sufficient condition than using a positive frame to define an OPD. 
We could allow some elements of the frame not to be positive, as long as the corresponding
environmental operator $\mbox{Tr}_S\left[(F^\dag_\alpha\otimes\mathbb{1}_E) \rho_{SE}\right]$ is equal to the null operator, and then also the corresponding $\omega_\alpha$ as in Eq.(\ref{eq:oma1}) is equal to zero. 
More in general, it might be useful to consider an OPD where the frame is chosen according
to the specific initial state (or set of initial states) at hand, and possibly even according to the 
global unitary evolution, 
so as to simplify the dynamical representation in Eq.(\ref{eq:opdt}) as much as possible.
We leave the study of how to optimize the choice of the frame in the sense now indicated
for future investigation.

\section{Existence and cost of the decomposition}\label{sec:eacc}

\subsection{Positive decomposition in finite and infinite-dimensional systems}
 \label{sec:posit-decomp-finite}
We introduce now two different constructions of an OPD valid for any global state $\rho_{SE}$; 
the first one is referred to a finite
dimensional Hilbert space $\mathcal{H}_S$ and relies on the definition of a positive frame, which generalizes straightforwardly the
frame for qubit systems introduced in \cite{Silva2019}. The second OPD, instead, includes 
fully general Hilbert spaces $\mathcal{H}_S$ and is defined in terms
of a family of positive operators that, in the infinite dimensional case, do not constitute a frame 
but still allow us to
exploit a reconstruction formula as in Eq.(\ref{eq:decbip}).

Let us then first consider a $d$-dimensional $\mathcal{H}_S$, with $d<\infty$, and the
$d^2$ generalized
Pauli matrices (also known as generalized Gell-Mann matrices) \cite{Kimura2003}
\begin{align}
\label{eq:CAP_d_Pauli}
\mathfrak{f}_{kj}^{(d)}&=\frac{1}{\sqrt{2}}
\begin{cases}
\ket{k}\bra{j} + \ket{j}\bra{k} \space ; \text{ for } k>j \\
-i\Big(\ket{j}\bra{k} - \ket{k}\bra{j} \Big) \space ; \text{ for } k<j
\end{cases} \nonumber\\
\mathfrak{h}_{k}^{(d)}&=
\begin{cases}
\frac{1}{\sqrt{d}} \mathbb{1}_d ; \text{ for } k=1 \\
\mathfrak{h}_{k}^{(d-1)}\oplus 0  ; \text{ for } k=2,\dots,d-1\\
\frac{1}{\sqrt{d(d-1)}} \big[\mathbb{1}_{(d-1)} \oplus (1-d)\big] ; \text{ for } k=d 
\end{cases},
\end{align}
where $\left\{\ket{k}\right\}_{k=1,\ldots d}$ is an orthonormal basis of $\mathcal{H}_S$
and $\oplus$ denotes the matrix direct sum; these matrices are Hermitian 
and traceless, apart from $\mathfrak{h}_{1}^{(d)}$ with trace $\sqrt{d}$,
and they correspond to the Pauli and Gell-Mann matrices
for $d=2$ and $d=3$, respectively.
A positive frame is thus defined by setting (with a mapping of the 
index $\alpha$ to the couple $(k,j)$)
\begin{align}
\label{eq:pauli_frame}
F_{11}^{(d)}&= \mathfrak{h}_{1}^{(d)}; \notag\\
F_{kk}^{(d)}&=\sqrt{d}\sqrt{\frac{k-1}{k}} \mathfrak{h}_{1}^{(d)} +\mathfrak{h}_{k}^{(d)};  \text{ for } k=2,...,d \notag\\
F_{kj}^{(d)}&={\frac{1}{\sqrt{2}}} \mathbb{1}_d+\mathfrak{f}_{kj}^{(d)};   \text{ for } k\neq j=1,...,d. 
\end{align}
The positivity of the frame elements
in Eq.(\ref{eq:pauli_frame}) directly follows from the fact that the minimum eigenvalue
of $\mathfrak{h}_{k}^{(d)}$ is $-\sqrt{(k-1)/k}$, while $\sqrt{d}\mathfrak{h}_{1}^{(d)}$
is equal to the identity matrix,
and the minimum eigenvalue of $\mathfrak{f}_{kj}^{(d)}$
is $-1/\sqrt{2}$. 
The inverse of the frame map provides us with the canonical dual frame, 
see Eq.(\ref{eq:indu}),
\begin{align}
\label{eq:pauli_frame_dual}
D_{11}^{(d)}&= \mathfrak{h}_{1}^{(d)} - \sqrt{d}\sum_{k=2}^{d}  \sqrt{\frac{k-1}{k}}\mathfrak{h}_{k}^{(d)} -\sqrt{\frac{d}{2} } \sum_{j\neq k}  \mathfrak{f}_{kj}^{(d)};\notag\\
D_{kk}^{(d)}&=\mathfrak{h}_{k}^{(d)};  \text{ for } k=2,...,d \notag\\
D_{kj}^{(d)}&=\mathfrak{f}_{kj}^{(d)};   \text{ for } k\neq j=1,...,d.
\end{align}
We will then call \emph{Pauli-OPD} the decomposition in Eq.(\ref{eq:opd}) fixed by 
the positive frame in Eq.(\ref{eq:pauli_frame}) and its dual in Eq.(\ref{eq:pauli_frame_dual}),
see also Eqs.(\ref{eq:oma1}) and (\ref{eq:omarhoa}).
Importantly, we note that since $D_{11}^{(d)}$ 
is the only element of the dual frame with non-vanishing trace, Eq.(\ref{eq:opd})
implies that $\rho_E=\Tr_S[\rho_{SE}]$ coincides with the first environmental statistical operator
appearing in the Pauli-OPD, $\rho_E=\rho_1$.

Let us now consider a possibly infinite dimensional Hilbert space $\mathcal{H}_S$
and an orthonormal basis $\left\{\ket{k}\right\}_{k=1,\ldots, d; k\in \mathbbm{N}}$
(where the first set of values of $k$ refers to a finite $d$-dimensional
Hilbert space and the second to an infinite dimensional one; to simplify the notation, from now on
the range of values of $k$ will be implied).
We start by defining
\begin{align}\label{eq:infinite_basis_HS}
\mathfrak{b}_{kj}&=\frac{1}{\sqrt{2}}
\begin{cases}
\ket{k}\bra{j} + \ket{j}\bra{k} \space ; \text{ for } k>j \\
-i\Big(\ket{j}\bra{k} - \ket{k}\bra{j} \Big) \space ; \text{ for } k<j
\end{cases} \\
\mathfrak{b}_{kk}&= \ket{k}\bra{k}, \nonumber
\end{align}
that is an orthonormal basis of Hermitian operators in $\hs$;
note that in $d$ dimensions this basis differs from the basis of generalized Pauli operators only in its diagonal elements.
Using a basis of $\hse$ made of tensor product operators
between the elements of, respectively, the basis in Eq.(\ref{eq:infinite_basis_HS}) and a
basis of $\he$ (analogously to what has been done to derive Eq.(\ref{eq:decbip})), we can decompose
any $O_{SE}\in \hse$ as
\begin{equation}\label{eq:ose4}
O_{SE} = \sum_{k j} \mathfrak{b}_{k j} \otimes \mbox{Tr}_S\left[(\mathfrak{b}_{k j}\otimes \mathbb{1}_E) O_{SE}\right].
\end{equation}
If we now consider 2 families of operators
$P_{k j}=\sum_{k' j'} M^{k' j'}_{k j} \mathfrak{b}_{k' j'}$ and
$Q_{k j}=\sum_{k' j'} N^{k' j'}_{k j} \mathfrak{b}_{k'j'}$
such that the corresponding real coefficients satisfy
\begin{equation}\label{eq:cons3}
\sum_{k'' j''} M^{k j}_{k'' j''}N^{k' j'}_{k'' j''} = \delta_{k k'}\delta_{j j'},
\end{equation} 
Eq.(\ref{eq:ose4}) is equivalent to
\begin{equation}\label{eq:ose5}
O_{SE} = \sum_{k j} Q_{k j} \otimes \mbox{Tr}_S\left[(P_{k j}\otimes \mathbb{1}_E) O_{SE}\right];
\end{equation}
indeed, if the operators $P_{kj}$ are positive, with the same reasoning as the one at the end of 
Sec.\ref{sec:frame-theory-posit}, we can conclude that the previous relation provides any 
$\rho_{SE}\in \hse$ with a valid OPD.
Our choice for the positive operators, aimed at reproducing the non-diagonal elements of Eq.~(\ref{eq:pauli_frame}) and using simple diagonal operators, 
is the following:
\begin{align}
\label{eq:induced_pos_family}
	P_{kk}&=\ket{k}\bra{k}; \notag\\
	P_{kj}&=\frac{1}{\sqrt{2}} \mathbb{1} + \mathfrak{b}_{kj}; \text{ for } k\neq j.
\end{align}
Using Eq.(\ref{eq:cons3}), we can thus complete
the decomposition in Eq.(\ref{eq:ose5}) with the family of operators
\begin{align}
\label{eq:induced_dual_family}
	Q_{kk}&=\ket{k}\bra{k} - \frac{1}{\sqrt{2}} \sum_{k\neq j}\mathfrak{b}_{kj}; \notag\\
	Q_{kj}&=\mathfrak{b}_{kj}; \text{ for } k\neq j.
\end{align}
We will call the OPD fixed by Eqs.(\ref{eq:ose5})-(\ref{eq:induced_dual_family}) \textit{basis-induced OPD}, to stress its connection with the initial choice of
the basis $\left\{\ket{k}\right\}$ of $\mathcal{H}_S$. Such an OPD can be used both for finite and infinite
dimensional systems, but we stress that only in the former case the family of operators
defined in Eq.(\ref{eq:induced_pos_family}) is actually a frame.
In the infinite dimensional case, in fact, the upper bound in Eq.(\ref{eq:frcond}) does not hold,
as, e.g., if we consider the operator $A=\mathfrak{b}_{kk}=\ket{k}\bra{k}$ we find the divergent series 
$$
	\sum_{k'j'} |\Tr_S[\mathfrak{b}_{kk} P_{k'j'}] |^2 = 1+ \sum_{k'\neq j'} \frac{1}{2}.
$$
Thus, the basis-induced OPD shows that, strictly speaking, frames are not necessary
to define a proper OPD,
the key point rather being the definition of a reconstruction formula as in Eq.(\ref{eq:decbip}),
or, equivalently, Eq.(\ref{eq:ose5}) via a family of positive open-system operators ($F_\alpha$
or $P_{kj}$).
 
 \subsection{Cost of the decomposition and Schmidt rank of the initial global state}
 \label{sec:cost-decomp}
 Moving back to a finite-dimensional Hilbert space $\mathcal{H}_S$ with dimension $d$,
 both the Pauli and the basis-induced OPD allow us to express any global state
by means of an OPD with $d^2$ terms. On the other hand, assigned a certain state $\rho_{SE}$,
it might well be that it is possible to write it equivalently via an OPD with a lower number of terms. 
Recall that we denote with $\mathfrak{N}$ the minimal number of terms appearing in the OPD of a given state 
$\rho_{SE}$, that is the cost
of the OPD of $\rho_{SE}$; moreover, since we deal now with finite-dimensional
Hilbert spaces, we will consider OPDs induced by positive frames.
Here, we show that the cost $\mathfrak{N}$ of the OPD of a given state $\rho_{SE}$ is 
equal to the Schmidt rank of $\rho_{SE}$.
 
As first step, we show that given a bipartite statistical operator $\rho_{SE}$, 
the cost $\mathfrak{N}$ of its OPD equals the number $\mathfrak{I}$ of linearly independent operators in the set of environmental states $\rho_{\alpha}$ associated with non-zero coefficients 
$\omega_\alpha$. Importantly, this number does not depend on the specific frame chosen to perform the decomposition and therefore the cost is a property of the operator itself. 
Hence, consider any positive frame $\left\{F_{\alpha}\right\}_{\alpha=1,\ldots \mathfrak{D}}$ of $\hs$, 
with $\mathfrak{D}$
elements ($\mathfrak{D}\geq d^2$ as the frame spans $\hs$), along with a dual frame
$\left\{D_{\alpha}\right\}_{\alpha=1,\ldots \mathfrak{D}}$. 
As shown Sec.\ref{sec:frame-theory-posit}, given any bipartite statistical operator
$\rho_{SE}$, we can write its OPD with respect to this frame as
$$
\rho_{SE}= \sum_{\alpha=1}^{N} \omega_{\alpha} D_{\alpha}\otimes \rho_{\alpha},
$$
where $\rho_{\alpha}$
are environmental statistical operators defined via Eq.(\ref{eq:omarhoa}) and we have ordered 
the frame elements such that the first $N\leq \mathfrak{D}$ coefficients $\omega_\alpha$ defined as in Eq.(\ref{eq:oma1}) are strictly positive, while the last $\mathfrak{D}-N$ are equal to zero. 
Now, let us denote as $\mathfrak{I}$ the number of linearly independent $\rho_{\alpha}$
in the set $\left\{\rho_{\alpha}\right\}_{\alpha=1,\ldots N}$, which indeed coincides with the number of linearly independent operators in the set 
$\{\omega_{\alpha}\rho_{\alpha}\}_{\alpha=1,\ldots \mathfrak{D}}$; if $N>\mathfrak{I}$,
we can write $\rho_N=\sum_{\alpha=1}^{N-1} c_\alpha \rho_\alpha$ for some coefficients $c_\alpha \in\mathbbm{R}$, from which it follows that $\rho_{SE}$ is equivalently represented by
$$
\rho_{SE}=\sum_{\alpha=1}^{N-1} \omega_{\alpha} \overline{D}_{\alpha} \otimes \rho_{\alpha},
$$
where we introduced the new frame
\begin{align}
\overline{D}_{\alpha}&= D_{\alpha}+\frac{\omega_{N}}{\omega_{\alpha}} c_\alpha D_{N} \text{  ; for } 1\leq \alpha\leq N-1 \notag\\
\overline{D}_{\alpha}&= D_{\alpha} \text{  ; for } N\leq \alpha\leq \mathfrak{D}.
\end{align}
Correspondingly, the duality relation is preserved if we also define
\begin{align}
\overline{F}_{\alpha}&= F_{\alpha} \text{  ; for } 1\leq \alpha\leq N-1 \notag\\
\overline{F}_{N}&= F_{N} - \sum_{\alpha=1}^{N-1} \frac{\omega_{N}}{\omega_{\alpha}} c_\alpha F_\alpha \notag\\
\overline{F}_{\alpha}&= F_{\alpha} \text{  ; for } N+1\leq \alpha\leq \mathfrak{D}.
\end{align}
This frame is not necessarily positive since $\overline{F}_N$ may have negative eigenvalues,
but this does not affect the resulting OPD, as 
\begin{equation}\label{eq:extra1}
\Tr_S\big[(\overline{F}_{N}\otimes \mathbb{1}_E)\rho_{SE}\big]=0;
\end{equation} compare with the discussion
at the end of Sec.\ref{sec:frame-theory-posit}.
Indeed, this procedure can be repeated if also $\{\rho_{\alpha}\}_{\alpha=1,\dots,N-1}$ are linearly dependent, i.e., $N-1>\mathfrak{I}$, and so on: if there are $\mathfrak{I}$ linear independent operators in 
$\{\rho_{\alpha}\}_{\alpha=1,\dots,N}$, it is always possible to write an OPD with only $\mathfrak{I}$ terms.
Crucially, no further reduction is possible as can be shown by reductio ad absurdum.
Suppose, in fact, that one can 
construct a different OPD with $\mathfrak{I}'<\mathfrak{I}$ terms
$\rho_{SE}=\sum_{\beta=1}^{\mathfrak{I}'} \lambda_{\beta} Q_{\beta} \otimes \eta_{\beta}$,
starting from a possibly different frame of operators 
$\{Q_{\beta}\}_{\beta=1,\ldots, \mathfrak{D}'}$, dual to a positive frame 
$\{P_{\beta}\}_{\beta=1,\ldots, \mathfrak{D}'}$. 
For what seen above, the set of statistical operators $\{\eta_{\beta}\}_{\beta=1,\dots,\mathfrak{I}'}$ 
can be taken linearly independent and 
$\lambda_{\beta} = \Tr_{SE}\left[(P_{\beta}\otimes \mathbb{1}_E)\rho_{SE}\right]=0$ 
for $\beta>\mathfrak{I}'$, see Eq.(\ref{eq:extra1}), without loss of generality. But then using
the decomposition of $F_{\alpha}$ on the frame $\{P_{\beta}\}_{\beta=1,\ldots, \mathfrak{D}'}$,
$F_{\alpha}=\sum_{\beta=1}^{\mathfrak{D}'} q_{\beta \alpha} P_{\beta}$,
we obtain
\begin{align}
\omega_{\alpha}\rho_{\alpha}&= \Tr_S\big[(F_{\alpha}\otimes \mathbb{1}_E)\rho_{SE}\big] 
=\sum_{\beta=1}^{\mathfrak{D}'} q_{\beta \alpha} \lambda_{\beta}\eta_{\beta} =\sum_{\beta=1}^{\mathfrak{I}'} q_{\beta \alpha} \lambda_{\beta}\eta_{\beta}.
\end{align}
This equation states that the set of linearly independent vectors $\{\omega_{\alpha}\rho_{\alpha}\}_{\alpha=1,\dots,\mathfrak{I}}$ is generated by a family $\{\lambda_{\beta}\eta_{\beta}\}_{\beta=1,\dots,\mathfrak{I}'}$ with $\mathfrak{I}'<\mathfrak{I}$, which is a contradiction.
We can thus conclude that, starting from any OPD, it is always possible to reduce the number of terms appearing in it to the number $\mathfrak{I}$ of independent operators in the set 
$\{\omega_{\alpha}\rho_{\alpha}\}_{\alpha=1,\ldots \mathfrak{D}}$, but not more, i.e., 
$\mathfrak{N}=\mathfrak{I}$; note that the reasoning used in the reductio ad absurdum also implies that
$\mathfrak{I}$ does not depend on the specific frame used to define the OPD of $\rho_{SE}$.


The connection between the cost $\mathfrak{N}$ of an OPD and the number $\mathfrak{I}$ of linearly independent environmental states appearing in the OPD directly leads us to link 
$\mathfrak{N}$ with the Schmidt rank.
Any bipartite state $\rho_{SE}\in \mathcal{S}(\mathcal{H}_S \otimes \mathcal{H}_E)$
can be associated with a Schmidt decomposition, which reads: 
\begin{equation}
\label{eq:Appendix_Schmidt_dec_op}
\rho_{SE}=\sum_{k=1}^{\mathfrak{R}} \lambda_k G_k^S \otimes G_k^E,
\end{equation}
with $\lambda_k>0$ and $\{G_k^{S,E}\}_{k=1,\ldots \mathfrak{R}}$ orthonormal Hermitian operators
on $\mathcal{H}_S, \mathcal{H}_E$; $\mathfrak{R}$ is the \textit{Schmidt rank} of $\rho_{SE}$
and
$\mathfrak{R} \leq d^2$; importantly, the Schmidt rank is directly related with 
the amount of correlations in $\rho_{SE}$ \cite{Cariello2014a,Cariello2015a}. 
Now, from Eq. (\ref{eq:Appendix_Schmidt_dec_op}) it follows
\begin{equation}\label{eq:gke}
\lambda_k G_k^E= \Tr_S\big[(G_k^S\otimes \mathbb{1}_E)\rho_{SE}\big] 
\end{equation}
and that the operators $\{\lambda_k G_k^E\}_{k=1,\dots,\mathfrak{R}}$ are linearly independent.
For an OPD, see Eq.(\ref{eq:opd}), we have the similar relation Eq.~(\ref{eq:omarhoa}) and we can always take into
account a minimal OPD of $\rho_{SE}$ 
such that the $\mathfrak{N}$ operators 
$\{\omega_{\alpha}\rho_{\alpha}\}_{\alpha=1,\dots,\mathfrak{N}}$ are linearly independent, 
as discussed above. 
In addition, we consider, respectively, an orthonormal basis $\{G_k^S\}_{k=1,\ldots, d^2}$ of $\hs$ whose first $\mathfrak{R}$
elements coincide with the open-system Schmidt operators in Eq.(\ref{eq:Appendix_Schmidt_dec_op}),
also implying $\Tr_S\big[(G_k^S\otimes \mathbb{1}_E)\rho_{SE}\big] =0$
for $k=\mathfrak{R}+1,\ldots, d^2$,
and a complete frame $\{F_{\alpha}\}_{\alpha=1,\ldots,\mathfrak{D}}$ whose first $\mathfrak{N}$ elements coincide with those defining the minimal OPD, while the others 
satisfy $\Tr_S\big[(F_\alpha\otimes \mathbb{1}_E)\rho_{SE}\big] =0$
for $\alpha=\mathfrak{N}+1,\ldots, \mathfrak{D}$, see Eq.(\ref{eq:extra1}). 
Exploiting the decompositions,
$
G_k^S=\sum_{\alpha=1}^{\mathfrak{D}} g_{k \alpha} F_{\alpha}$
and $F_{\alpha}=\sum_{k=1}^{d^2} f_{\alpha k} G_k^S$,
we thus obtain from Eqs.~(\ref{eq:omarhoa}) and (\ref{eq:gke})
\begin{align}
\lambda_k G_k^E&=\sum_{\alpha=1}^{\mathfrak{N}} g_{k \alpha} \omega_{\alpha}\rho_{\alpha}  ; \notag\\
\omega_{\alpha}\rho_{\alpha}&=\sum_{k=1}^{\mathfrak{R}} f_{\alpha k} \lambda_k G_k^E .
\end{align}
Hence, since the linearly independent sets $\{\lambda_k G_k^E\}_{k=1,\dots,\mathfrak{R}}$ and $\{\omega_{\alpha}\rho_{\alpha}\}_{\alpha=1,\dots,\mathfrak{N}}$ generate each other, they have the same number of elements, i.e., $\mathfrak{N}=\mathfrak{R}$.
We thus conclude that the cost $\mathfrak{N}$ of the OPD coincides with the Schmidt rank of the global
statistical operator $\rho_{SE}$, in this way generalizing to mixed states the link between Schmidt rank and OPD discussed in \cite{Silva2019} for pure global states.

\section{Case study: combination of semigroup maps}\label{sec:csc}

We now provide an explicit example of an open-system evolution fixed via a family of CPTP maps
$\Phi_{\alpha}(t)$
according to Eq.(\ref{eq:opdt}). 
In Ref.\cite{Trevisan2021a} we exploited the possible relevance of OPD in order to obtain a perturbative expansion of a microscopically motivated open system dynamics in the presence of initial correlations. We will here take a complementary phenomenological approach, exploring whether convenient choices of the CPTP maps $\Phi_{\alpha}(t)$ can lead to a well defined description of the system evolution.
In particular, we take the maps $\Phi_{\alpha}(t)$ to be CPTP semigroups, i.e.,
\begin{equation}\label{eq:phisemi}
\Phi_{\alpha}(t) = e^{\mathcal{L}_\alpha t},
\end{equation}
where $\mathcal{L}_\alpha$ are generators in the
GKLS form \cite{Lindblad1976,Gorini1976}
\begin{equation}
\mathcal{L}_\alpha[\rho] = \sum_{j=1}^{d^2-1}\gamma_{\alpha,j}\left( 
L_{\alpha, j} \rho L^\dag_{\alpha, j}
-\frac{1}{2}\left\{L^\dag_{\alpha,j } L_{\alpha,j}, \rho\right\}\right),
\end{equation}
with $\gamma_{\alpha,j}\geq 0$ and $L_{\alpha,j}$ linear independent linear operators on $\mathcal{H}_S$, 
and we recall that $d$ is the finite dimension of the open-system
Hilbert space $\mathcal{H}_S$.

The idea of combining a family of semigroup maps to go beyond the description of Markovian
dynamics has been used by Chru{\'s}ci{\'n}ski and Kossakowski in several pioneering 
works \cite{Chruscinski2010,Chruscinski2011,Chruscinski2012},
aiming at the identification of well-defined integro-differential master equations accounting for 
non-Markovian dynamics. 
In \cite{Breuer2007a} a system of GKSL evolutions was used to take into account statistical correlations between system and environment leading to a non-Markovian dynamics within a projection operator approach. Here, by using Eq.(\ref{eq:opdt}) to connect the initial reduced state
with its value at a generic time, we explore whether a family of semigroups can be exploited 
in the context
of open-system dynamics in the presence of initial correlations with the environment.
Indeed, while every $e^{\mathcal{L}_\alpha t}$ is CPTP,
so that also a convex combination of these maps would be CPTP, we are now applying
each $e^{\mathcal{L}_\alpha t}$ to a distinct element of the dual frame $D_\alpha$,
which needs not to be positive. As a consequence, it is a-priori not guaranteed that even though
we start from an initial positive state $\rho_S$ in Eq.(\ref{eq:inits}),
the state at time $t$ fixed by Eq.(\ref{eq:opdt}) will be positive, too.
In the following, 
we show that it is actually possible to identify a proper set of initial states that are mapped
into states at any time $t$, for a relevant class of two-level system dynamics.
Let us stress that, as common in phenomenological approaches, the fact that we can
introduce a well-defined evolution on open-system states does not mean by itself that the evolution
can be derived from a full microscopic model. Within the formalism exploited here, 
the existence of a microscopic model would consist in the presence of
initial global states and a global unitary evolution such that the action of the maps
$\Phi_{\alpha}(t)$ could be expressed as in Eq.(\ref{eq:phialpha}); 
whether this is the case is a challenging question,
which we leave for future investigation.

\subsection{Pauli channels}

We consider a two-level open quantum system and we further exploit
the positive frame induced by the Pauli basis of operators, introduced in 
Sec.\ref{sec:posit-decomp-finite}.
In particular, the dual-frame operators $D_\alpha$ are as in Eq.(\ref{eq:pauli_frame_dual}),
which for the case $d=2$ we are dealing with 
simply read
\begin{align}
D_0&=\frac{1}{\sqrt{2}}\left(\sigma_0 - \sum\limits_{\alpha=1}^3 \sigma_{\alpha}\right), & \nonumber\\
D_{\alpha}&=\frac{1}{\sqrt{2}}\sigma_{\alpha}, && \text{for }{\alpha}=1,2,3,
\end{align}
where $\sigma_0$ is the identity, the $\sigma_\alpha$, $\alpha=1,2,3$ are the usual Pauli matrices,
and we reordered the frame indices as $\alpha = 0, \ldots, 3$.
Given a reduced operator of the form (compare with Eq.(\ref{eq:inits}))
\begin{align} \label{eq:red}
\rho_S=\frac{1}{\sqrt{2}}\sum\limits_{\alpha=0}^3 \upsilon_\alpha D_\alpha,
\end{align}
where for the sake of convenience we have set $\upsilon_\alpha=\sqrt{2}\omega_\alpha$ with $\omega_\alpha$ as in Eq.(\ref{eq:inits}),
it has trace one if and only if
\begin{align}
\upsilon_0=1,
\end{align}
while its positivity is ensured by the validity of
\begin{align}\label{eq:cond1}
\left(1 - \upsilon_1\right)^2+\left(1 - \upsilon_2\right)^2+\left(1 - \upsilon_3\right)^2 \leq 1.
\end{align}
Hence, the initial positivity domain, that is the set of points $\{v_1,v_2,v_3\}$ for which $\rho_S$ as in Eq.(\ref{eq:red})
can be taken as a proper initial state,
is a ball with radius 1 and origin in the point $\{1,1,1\}$. 

Our goal is then to determine whether there is a subset of the initial positivity domain such that the state
$\rho_S(t)$ obtained via Eq.(\ref{eq:opdt}) is positive at any time $t$.
In particular, we take into account the maps corresponding to the so-called 
Pauli channels \cite{Vacchini2012a,Chruscinski2013}, whose generators are given by
\begin{equation}
\mathcal{L}_{\alpha}[\rho] = \sum_{j=1}^3 \gamma_{\alpha,j}(t)\left(\sigma_j \rho \sigma_j - \rho\right).
\end{equation}
When the rates $\gamma_{\alpha,j}(t)$ are non-negative and time independent each of this generator corresponds to a GKSL master equation, otherwise the dynamics can be highly non-Markovian also for a single $\mathcal{L}_{\alpha}$ \cite{Megier2017}.   
Here, as said, we restrict to semigroups; note that while the coefficients $\gamma_{\alpha,j}$ can be different for the different 
$\mathcal{L}_{\alpha}$s, the Lindblad operators are always the same, i.e., the Pauli matrices.
The corresponding CPTP maps take the form
\begin{align}
\phi_\alpha(t) [\rho]=\sum\limits_{j=0}^3 p_{\alpha,j}(t) \sigma_j\rho \sigma_j,
\end{align} 
where the coefficients $p_{\alpha,j}(t)$ are readily expressed by introducing 
the vectors $\vec p_{\alpha}(t)$, whose components are
$(\vec p_{\alpha}(t))_j = p_{\alpha,j}(t)$, 
so that one has
\begin{align}
\vec p_{\alpha}(t)=\frac{1}{4} \begin{pmatrix}
1 & 1 & 1 & 1\\
1 & 1 & -1 & -1 \\
1 & -1 & 1 & -1 \\
1 & -1 & -1 & 1 
\end{pmatrix}
\vec \lambda_{\alpha}(t)
\end{align}
where $(\vec \lambda_{\alpha}(t))_j=\lambda_{\alpha,j}(t)$ with
\begin{align}
\lambda_{\alpha,0} =1, &&\lambda_{\alpha,j} = e^{-2 (\gamma_{\alpha,k} + \gamma_{\alpha,m})t}, &&  j \neq k \neq m = 1,2,3,
\label{eq:gammas}
\end{align}
i.e., the exponential decays characterizing semigroup dynamics.
Note that the $\left\{\lambda_{\alpha,j}(t)\right\}_{j=0,\ldots,3}$ are eigenvalues of the map $\phi_\alpha(t)$
\begin{align}
\phi_\alpha(t) [\sigma_j]=\lambda_{\alpha,j}(t) \sigma_j,
\end{align}
while the coefficients $\left\{p_{\alpha,j}(t)\right\}_{j=0,\ldots,3}$ form a probability distribution, 
i.e. $p_{\alpha,j}(t) \geq 0$ and 
$\sum\limits_{j=0}^3 p_{\alpha,j}(t)=1$.   

For the sake of simplicity, we now restrict to two different semigroup
maps, one for the dual frame element $D_0$, which we denote as $\phi(t):=\phi_0(t)$,
and another for the other elements, which we denote as
$\tilde \phi(t):=\phi_1(t) = \phi_2(t) =\phi_3(t)$; accordingly all the parameters referred to 
the first (second) map will be indicated without (with) a tilde. 
The state at time $t$ obtained via Eq.(\ref{eq:opdt}) then reads
\begin{align}
\rho_S(t) = \frac{1}{2}\left( \opone + \left(\tilde\lambda_1(t)v_1 -\lambda_1(t)\right)\sigma_1+ \left(\tilde\lambda_2(t)v_2 -\lambda_2(t)\right)\sigma_2+ \left(\tilde\lambda_3(t)v_3 -\lambda_3(t)\right)\sigma_3
\right),
\end{align}
so that the positivity domain at a generic time is specified by the condition
\begin{align}\label{eq:cond2}
\left(\lambda_1(t) - \tilde\lambda_1(t)v_1\right)^2+\left(\lambda_2(t) - \tilde\lambda_2(t)v_2\right)^2+\left(\lambda_3(t) - \tilde\lambda_3(t)v_3\right)^2 \leq 1,
\end{align}
which defines the interior of an axis-aligned ellipsoid with semi-axes lengths 
$1/\tilde\lambda_1(t)$, $1/\tilde\lambda_2(t)$ and $1/\tilde\lambda_3(t)$, centered at the point $\{\lambda_1(t)/\tilde\lambda_1(t),\lambda_2(t)/\tilde\lambda_2(t),\lambda_3(t)/\tilde\lambda_3(t) \}$.

\subsection{Results}

Summarizing the previous discussion, our primary goal is thus to find the values 
$\{v_1,v_2,v_3\}$ that satisfy both Eq. \eqref{eq:cond1} and \eqref{eq:cond2}, where the former inequality defines the initial positivity domain, while the latter depends on the dynamical parameters $\lambda_\alpha(t)$ and $\tilde\lambda_\alpha(t)$.

To this aim, it is advantageous to decouple the quantities referred to, respectively, 
$\phi(t)$ and $\tilde \phi(t)$. 
Note that we can interpret Eq. \eqref{eq:cond2} as an equation of a
ball with the axes rescaled by $\tilde\lambda$s:   \begin{align}\label{eq:cond3}
\left(\lambda_1(t) - \tilde x(t)\right)^2+\left(\lambda_2(t) - \tilde y(t)\right)^2+\left(\lambda_3(t) - \tilde z(t)\right)^2 \leq 1.
\end{align}
We call the set of points $\{\tilde x(t), \tilde y(t), \tilde z(t)\}$ satisfying this inequality an evolved positivity ball, as for $t=0$ it gives the initial positivity domain as in Eq. \eqref{eq:cond1}.
On the other hand, Eq. \eqref{eq:cond1} after such rescaling takes the form
\begin{align}\label{eq:cond4}
\frac{1}{\tilde\lambda_1^2(t)}\left(\tilde\lambda_1(t) - \tilde x(t)\right)^2+\frac{1}{\tilde\lambda_2^2(t)}\left(\tilde\lambda_2(t) - \tilde y(t)\right)^2+  \frac{1}{\tilde\lambda_3^2(t)}\left(\tilde\lambda_3(t) - \tilde z(t)\right)^2 \leq 1,
\end{align}
and the set of all evolved initial points is an axis-aligned ellipsoid centered at $\{\tilde\lambda_1(t),\tilde\lambda_2(t),\tilde\lambda_3(t) \}$ with semi-axes lengths $\tilde\lambda_1(t)$, $\tilde\lambda_2(t)$ and $\tilde\lambda_3(t)$. 
Accordingly, the intersection of this ellipsoid with the evolved positivity ball with radius 1 and origin in $\{\lambda_1(t),\lambda_2(t),\lambda_3(t)\}$ gives the wanted set of $v$s. If the whole ellipsoid is contained within the evolved positivity ball at a given time $t$, all initial density operators $\rho_S$ stay positive at this point of time. By virtue of the rescaling introduced above, the form and location of the ellipsoid is only governed by the $\tilde\lambda$s, while the positivity ball evolves according to the $\lambda$s.

\begin{figure*}
 \includegraphics[width=0.33\textwidth]{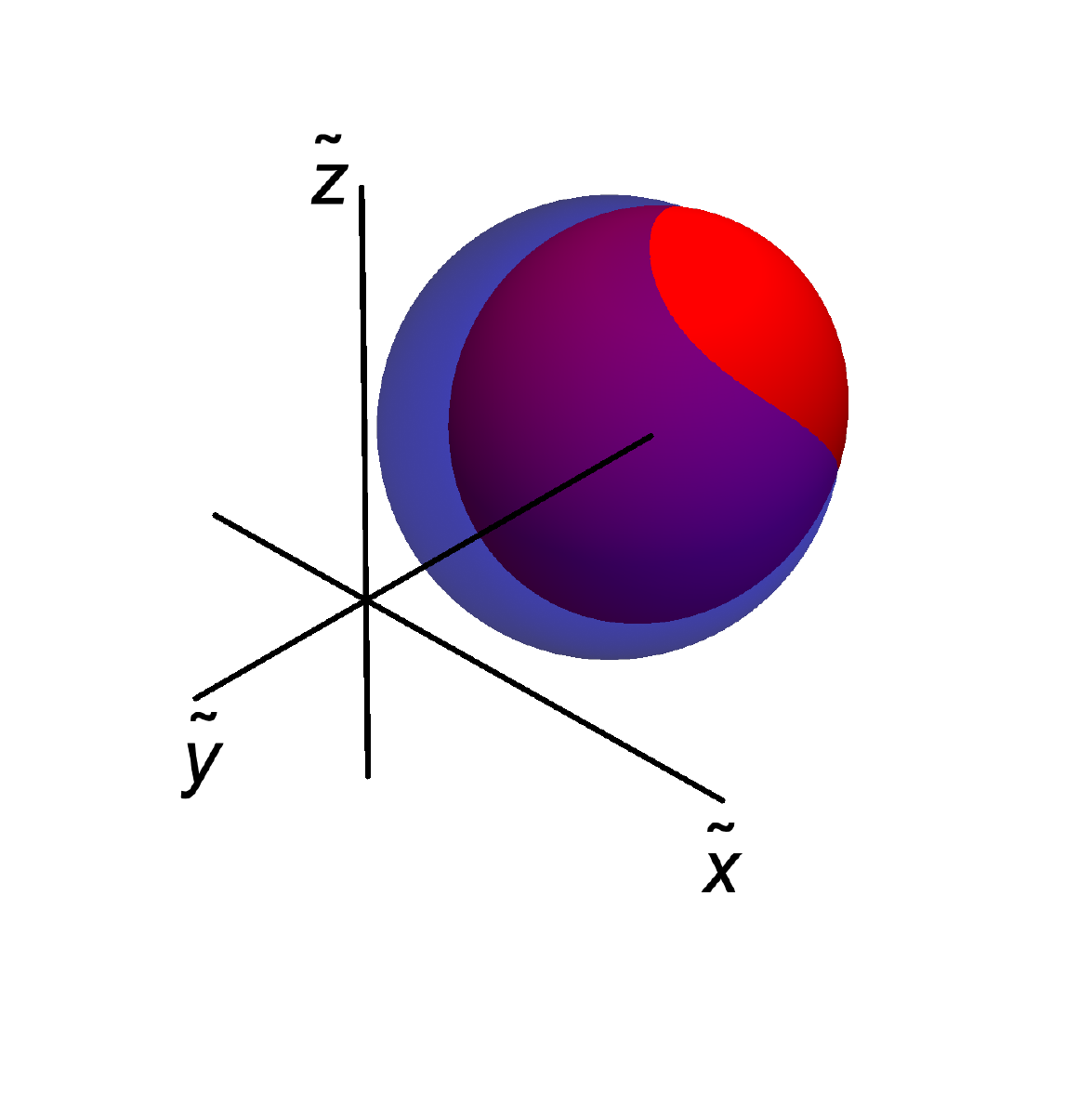}~~\includegraphics[width=0.33\textwidth]{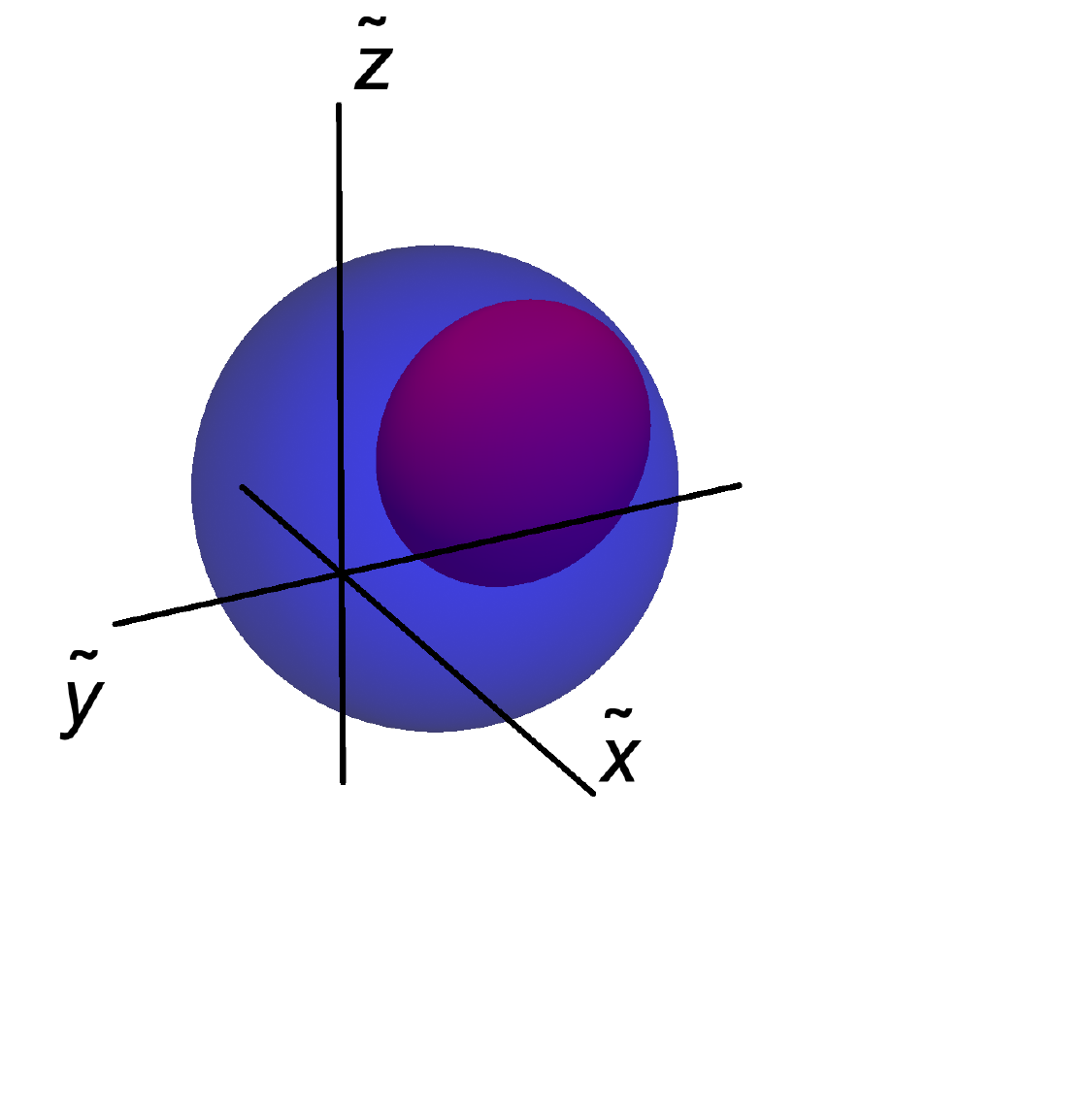}~~\includegraphics[width=0.33\textwidth]{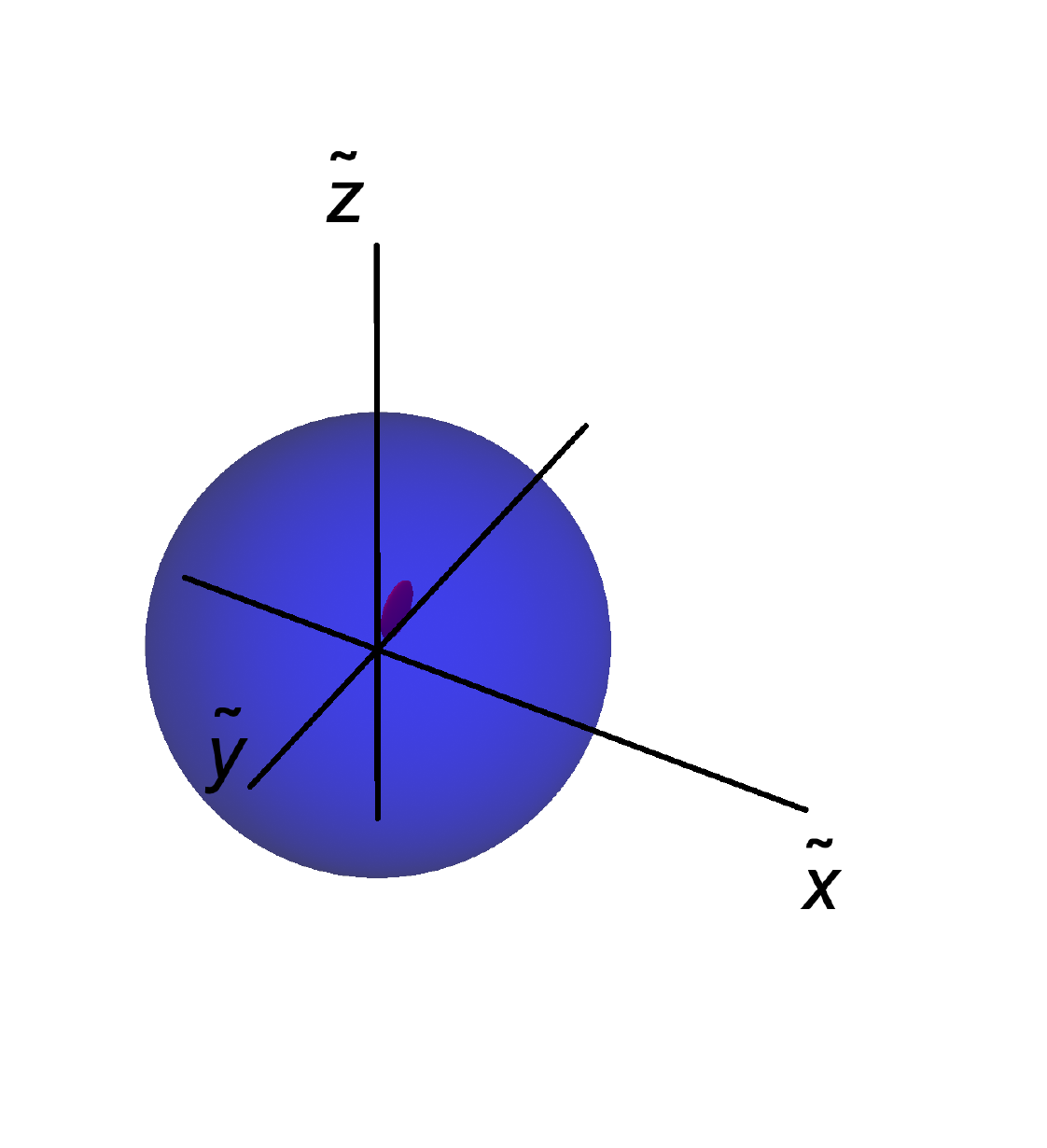}
\caption{
Example I: evolved initial points 
 (red) and the evolved positivity ball (blue). Left: $\gamma t = 1/10$, Middle: $\gamma t = 1/2$, Right: $\gamma t =2$.}
\label{fig:PC2}
\end{figure*}

\begin{figure*}
 \includegraphics[width=0.33\textwidth]{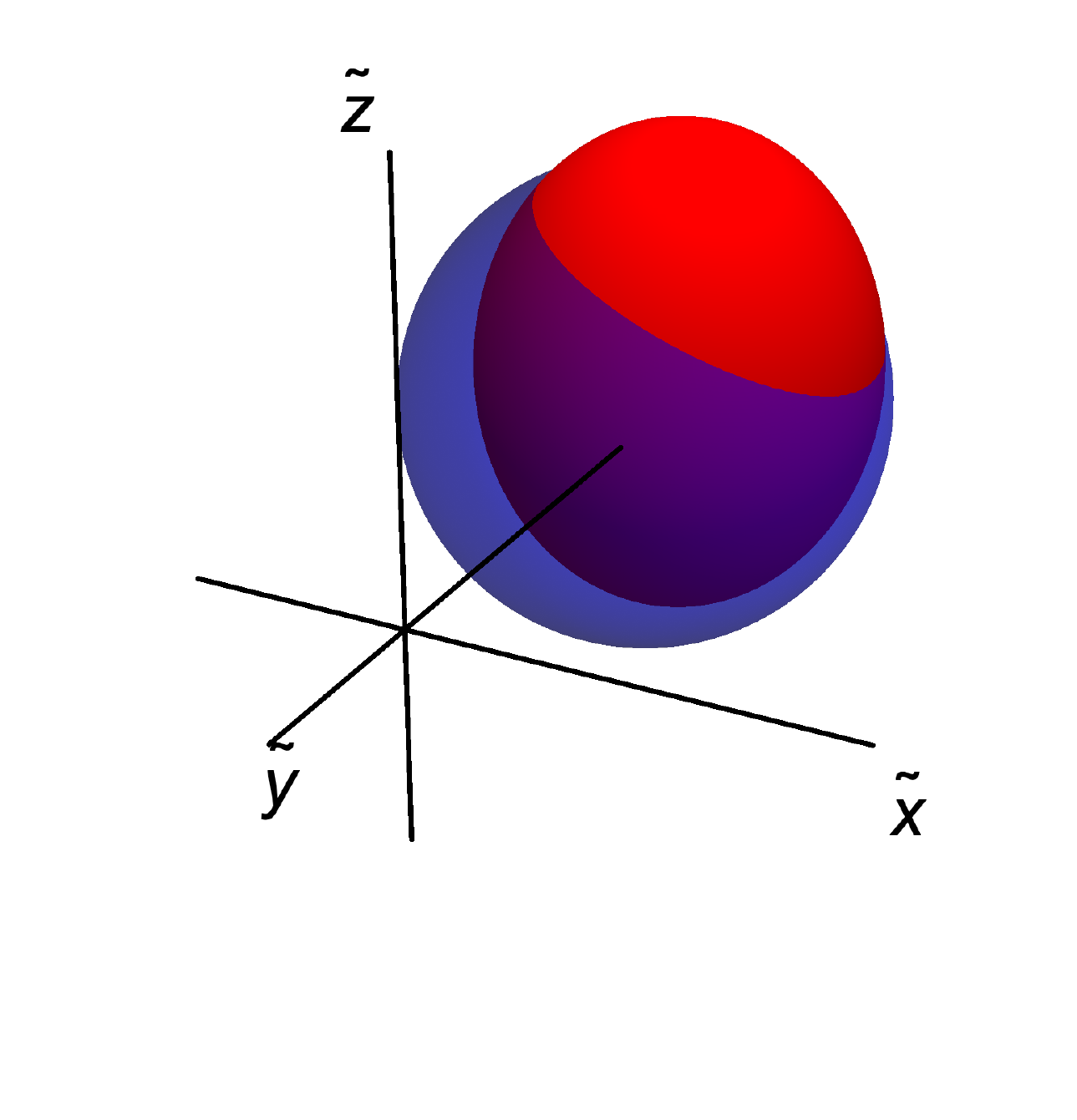}~~\includegraphics[width=0.33\textwidth]{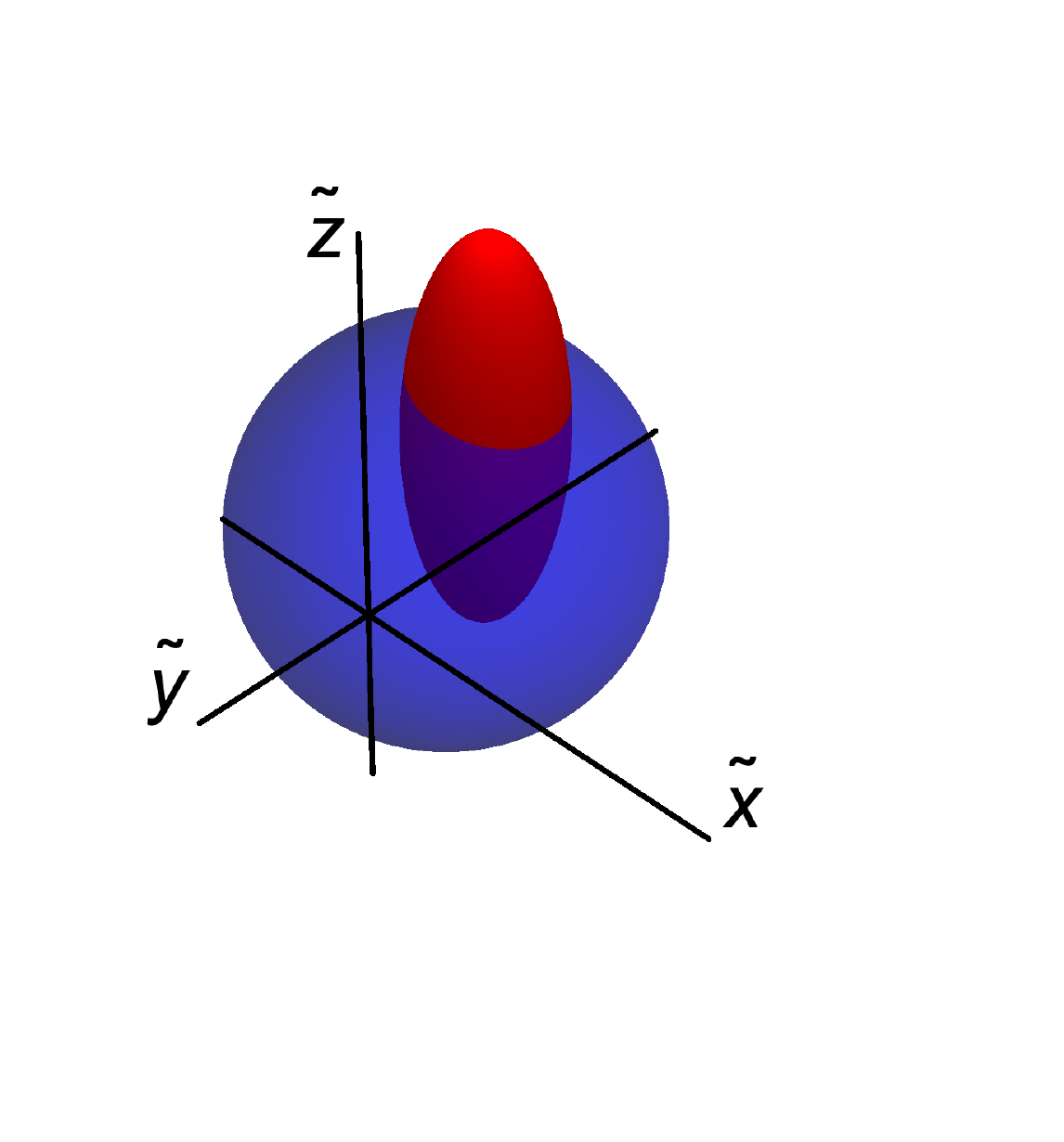}~~\includegraphics[width=0.33\textwidth]{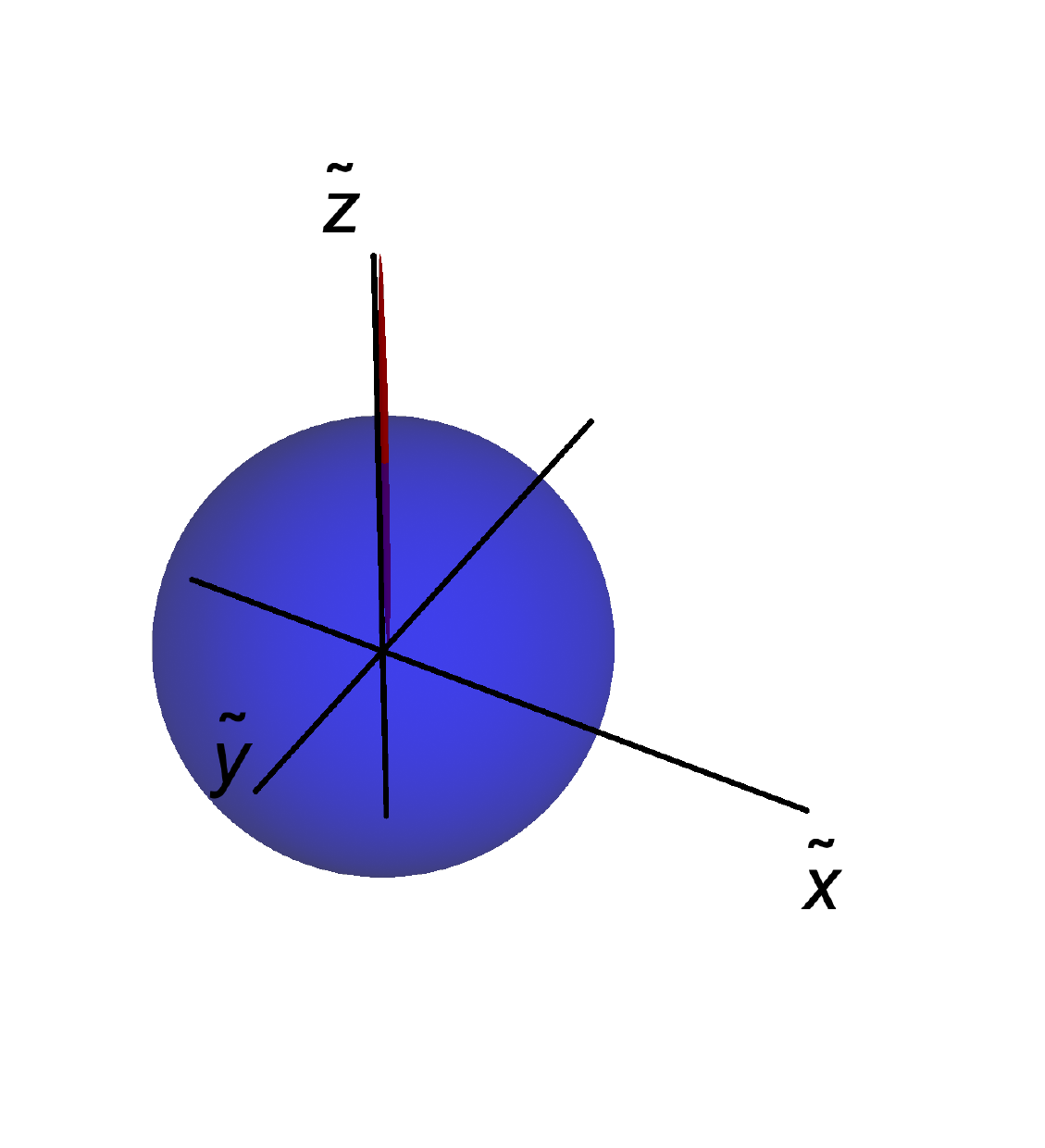}
\caption{Example II: evolved initial points (red) and the evolved positivity ball (blue). Left: $\gamma t = 1/10$, Middle: $\gamma t = 1/2$, Right: $\gamma t =2$.}
\label{fig:PC1}
\end{figure*}

In Fig.\ref{fig:PC2} and \ref{fig:PC1} we report the ellipsoid of evolved intial points defined by Eq.(\ref{eq:cond4}) (in red)
and the evolved positivity ball defined by Eq.(\ref{eq:cond3}) (in blue), for two different choices of the 
parameters defining the semigroup maps.

Example I: In a generic case all the $\lambda$s and $\tilde\lambda$s describe exponential decay, i.e. according to Eq.(\ref{eq:gammas}) at most one of the $\gamma$s and $\tilde\gamma$s is zero. In this case the centers of both the evolved positivity ball and the evolved ellipsoid converge to $\{0,0,0\}$. Additionally, the ellipsoid asymptotically shrinks to a point, while the size of the evolved positivity ball does not change. Hence, for large enough times the ellipsoid will be completely contained into the positivity ball. 
For intermediate times, a portion of the ellipsoid can leave the evolved positivity ball, as seen in Fig. \ref{fig:PC2} for the following choice of the parameters: $\gamma_1=\tilde\gamma_1=0$, $ \gamma_2=\gamma_3=2\tilde\gamma_2=2\tilde\gamma_3=\gamma$,
which correspond to
\begin{align}
\lambda_1(t)=e^{-4\gamma t},  && \lambda_2(t)=e^{-2\gamma t},  && \lambda_3(t)=&e^{-2\gamma t},\notag\\
\tilde\lambda_1(t)=e^{-2\gamma t}, && \tilde\lambda_2(t)=e^{-\gamma t},  && \tilde\lambda_3(t)=&e^{-\gamma t}.
\end{align}
The larger values of the $\gamma$s with respect to the $\tilde\gamma$s result in a quicker motion of the positivity ball than of the ellipsoid.  
Accordingly, some of the states in the initial positivity domain will anyhow lead to temporarily negative evolved matrix $\rho_S(t)$. 

Example II: If, on the other hand, two of the $\tilde\gamma$s equal zero, one can see a qualitatively different behaviour, as some of the choices of initial points lead to eternally negative evolved matrix $\rho_S(t)$. For example, 
for $\gamma_1=\tilde\gamma_1=\tilde\gamma_2=0$, $ \gamma_2=\gamma_3=\tilde\gamma_3=\gamma$,
which correspond to
\begin{align}
\lambda_1(t)=e^{-4\gamma t},  && \lambda_2(t)=e^{-2\gamma t},  && \lambda_3(t)=&e^{-2\gamma t}, \notag\\
\tilde\lambda_1(t)=e^{-2\gamma t}, && \tilde\lambda_2(t)=e^{-2\gamma t},  && \tilde\lambda_3(t)=&1,
\end{align}
the $z$-coordinate of the initial points is constant: $\{e^{-2\gamma t}v_1,e^{-2\gamma t}v_2,v_3\}$. Accordingly, the semi-axes length in this direction of the ellipsoid does not change in time and its center remains on the same x-y plane, see Fig. \ref{fig:PC1}. All of the choices of points with $v_3>1$ will at some point of the evolution leave the evolved positivity ball forever.

\section{Conclusion}\label{sec:con}
In this paper, we have considered an approach to describe the dynamics of open quantum systems
that allows one to take into account fully general initially correlated system-environment states.
The starting point is the OPD decomposition of the initial global state, which involves positive and normalized
operators on the environmental side and fixes the evolution of the open system through a set of time-dependent CPTP maps. 
Such a decomposition can be defined via a positive frame
on the open quantum system and it is in general highly non-unique.
In particular, we introduced two OPD decompositions, one that is based on the generalized Pauli matrices
and one directly on the basis elements of the open-system Hilbert space; notably, the latter
can be defined also for infinite-dimensional Hilbert spaces, thus showing how the OPD-based approach 
to open-system dynamics is not restricted to the finite-dimensional case.
In addition, we have demonstrated that the cost of the decomposition, that is the number of terms involved in it
and thus the number of resulting CPTP maps fixing the reduced dynamics, is always bounded by the Schmidt
rank of the initial global state. This is a particularly useful feature of the approach, since it implies
that the open-system evolution can be fixed by a number of equations, for example as in \cite{Trevisan2021a}, 
that is a direct expression of the amount of correlations between the open system and the environment and it is ultimately bounded
by the dimensionality of the open system.
Finally, we have studied the case where the CPTP maps defining the reduced dynamics along with the OPD decomposition
are semigroups fixed by the Gorini-Kossakowski-Lindblad-Sudarshan form.
In particular, in the case of qubit Pauli channels, we have derived explicit conditions identifying the positivity
domain defined by the initial reduced states that are mapped to proper states at any time 
and we have distinguished two qualitatively different regimes; one where all the states compatible
with a given initial decomposition are eventually mapped into proper states after a transient time interval,
and one where some states lead to eternally negative evolved matrices.

\section*{Acknowledgments}
\label{sec:acknowledgments}
All authors acknowledge support from UniMi, via Transition Grant H2020, PSR-2 2020 and PSR-2 2021. NM was funded by the Alexander von Humboldt Foundation in form of a Fedor-Lynen Fellowship and project ApresSF, supported by the National Science Centre under the QuantERA programme, funded by the European Union's Horizon 2020 research and innovation programme.

\bibliography{Bibliography}

\end{document}